# Managing Semantic Loss during Query Reformulation in Peer Data Management Systems

Yannis Delveroudis, Paraskevas V. Lekeas
*School of Electrical and Computer Engineering*
*National Technical University of Athens, Greece*
ydelver@gmail.com, plekeas@mail.ntua.gr

**Abstract**

*In this paper we deal with the notion of semantic loss in Peer Data Management Systems (PDMS) queries. We define such a notion and we give a mechanism that discovers semantic loss in a PDMS network. Next, we propose an algorithm that addresses the problem of restoring such a loss. Further evaluation of our proposed algorithm is an ongoing work.*

## 1. Introduction

In this paper we deal with the problem of semantic loss in PDMSs. A PDMS (Peer Data Management System) is an organized system of communicating nodes that participate actively in exchanging information [2]. Note that the word "semantics" is not used in its formal sense but refers in an informal way to the probable loss of potential answers to the query, because during its reformulation along several peers the composition of schema mappings do not capture all the relationships between two peers due to the lack of direct connection between them (see also 1.2 for a motivating example). As opposed to traditional data integration systems, these systems do not assume the existence of a mediated schema to which every node must map in order to share data. This fact enables PDMSs to be more extensible and scalable than data integration architectures, since they allow any user to contribute new data or schemas in a fully distributed manner. Also, the addition of new nodes in the system is easier, because they are able to choose to which existing peers they will map (usually the ones with the most similar schemas). Each peer uses its own local schema to pose queries and the existence of mappings between nodes enables the propagation of queries to the other peers of the network. Since the peers employ different schemas, the query gets reformulated during its propagation. Due to the dynamic structure of these systems (since peers are free to join and leave at will), queries often have to travel long paths in the network in order to reach useful peers (i.e., peers that satisfy the information needs of the inquiring peer). These reformulations many times introduce changes in the semantics of the original query, which affect negatively the quality of answers. We use the term "semantic loss" to refer to this problem-effect. Having in mind that a great number of queries are processed in large PDMSs, even small changes of semantics may affect dramatically the quality of answers in such networks.

The problem of semantic loss, although not directly studied, arises indirectly in the bibliography. For example, in the Piazza PDMS [4, 3], the queries lose information, due to differences between schemas of nodes. This happens due to semantic loss and affects the whole network, making it biased towards some query paths that preserve information and others that lose it.

Our contributions in this paper are the following:

- We define the notion of "semantic loss" of a query in a PDMS network.
- We propose a mechanism that tracks and restores such semantic losses.

This paper is organized as follows: In section 1.1 we refer to the related work, while in section 1.2 we present a motivating example that captures the problem of semantic loss. In section 2 we define the problem formally and propose an algorithm for solving it. Section 3 refers to some preliminary results of our experiments and finally section 4 concludes our work and discusses our further research.

### 1.1. Related work

To the best of our knowledge, the notion of semantic loss has not been treated by itself alone, but always indirectly in the context of the information loss problem. Examples of the information loss problem can be found in [1, 6]. Indirect references for the semantic loss exist in [3], where the authors propose that analyzing mapping networks for information loss is an interesting challenge to investigate.

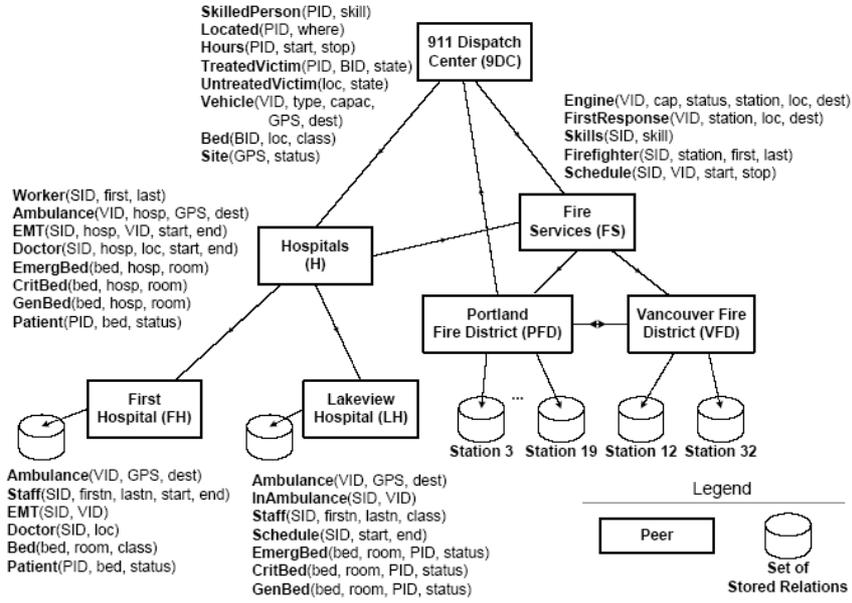

**Figure 1.** A PDMS for integrating data from emergency services. Arrows indicate semantic mappings.

### 1.2. Motivating example

Figure 1 illustrates a PDMS for emergency services (the system is a modified version of the one given in [4]). The SkilledPerson relation of the 9DC node can be expressed using a GAV (Global As View) mapping over the H schema, as follows:

9DC : SkilledPerson(PID, "Doctor") :-
 H : Doctor(SID, h, l, s, e)

9DC : SkilledPerson(PID, "EMT") :-
 H : EMT(SID, h, vid, s, e)

where PID = SID. Similarly, we can express the H relations as views over the LH relations. A part of this mapping is the following (note that we use again the GAV formalism):

H : Doctor(SID, "LH", "Portland", s, e) :-
 LH : Staff(SID, fn, ln, "Doctor"),
 LH : Schedule(SID, s, e)

H : EMT(SID, "LH", vid, s, e) :-
 LH : Staff(SID, fn, ln, "EMT"),
 LH : Schedule(SID, s, e),
 LH : InAmbulance(SID, vid)

Suppose now that a user at the 9DC peer wants information about all the skilled personnel in this area and thus he poses the following SQL query, using his own schema (since SQL is widely used to pose queries, when we refer to the user queries we use SQL instead of the conjunctive queries formalism):

$Q_1$: Select PID, skill
From SkilledPerson

In order for the query to retrieve data from the other peers of the system, it needs to be reformulated accordingly. Using the mapping between the 9DC and H schemas, the query is posed at the H node as a union:

$Q'_1$ : Select SID, "Doctor"
From Doctor
UNION
Select SID, "EMT"
From EMT

Note that the query has lost part of its semantics due to the reformulation (it has become more selective). Next, the query needs to propagate further and thus a second rewriting is necessary, using the mapping between the H and LH schemas.

Select SID, "Doctor"
From Staff, Schedule
Where class = "Doctor" and Staff.SID = Schedule.SID
UNION
Select SID, "EMT"
From Staff, Schedule, InAmbulance
Where class = "EMT" and Staff.SID = Schedule.SID
and Staff.SID = InAmbulance.SID

This query is even more restrictive than the original version due to the joins. Nevertheless, we could substitute the literals of the projection list (i.e., "Doctor" and "EMT") with the attribute *class* by virtue of the predicates class = "Doctor" in the first query and class = "EMT" in the second one and thus turn the query in a more flexible form, closer to its original version. Note that this is a subtle issue and is not guaranteed to always be semantically correct. In any case, we observe that the query has lost information and we need to find a way in order to avoid this loss. To make this more concrete, suppose that there was a direct mapping between the 9DC and LH peers like the following one:

9DC : SkilledPerson(PID, skill) :-
LH : Staff(SID, fn, ln, class)

where PID = SID and skill = class are taken as granted. Using this mapping, we could reformulate the original query over the LH schema in just one step as follows:

Select SID, class
From Staff

Comparing the two-step process with the hypothetical one-step translation above, it is obvious that the query loses information, since it asks about doctors and EMTs instead of all the skilled personnel. A direction for solving this problem is to consider adding some metadata to the reformulated query that capture the lost semantics at each step. This would enable us to recover it later, if there exists such a chance, although this metadata should be globally understandable (through a global ontology for instance), which contradicts with the peer-to-peer paradigm. Another idea to consider is adding metadata about the lost information in the mappings (and extending accordingly the reformulation algorithm), but that would transfer the problem of heterogeneity to the management of metadata.

## 2. A semantics-preserving mechanism

As we have observed in the motivating example there exist transformations between two peers that do not preserve the semantics. The first problem we must address in this context is the following: How can we track a loss of semantics in a PDMS network? Next, we would have to answer the question: How can we replace such a loss of semantics in a PDMS network? To the best of our knowledge there exists no previous work that considers these problems. In what follows we try to tackle these two problems by proposing an algorithm that tracks and replaces such losses. In section 2.1 we describe a mechanism that tracks and restores such semantic losses. In section 2.2 we describe how the restored semantics affects the whole PDMS network.

### 2.1. Tracking and replacing losses of semantics between two peers in a PDMS network

In order to track a loss of semantics in a PDMS network suppose we have the following setting. Given a query $Q$ from peer $P_1$ and its reformulated version $Q'$ on peer $P_2$, how can we compare them semantically, considering that they are formulated over different schemas? To answer this question, we propose a feedback mechanism that translates the rewritten query back to its originating peer (we denote this query as $Q''$). This casts the problem to a simpler one that is easier to reason about, since the two queries $Q$ and $Q''$ are now over the same schema (i.e., the schema of $P_1$).

**Definition 2.1 (Semantic Loss)** Let $Q$ be a query on peer $P_1$ and its reformulated version $Q'$ on peer $P_2$. Let also $Q''$ be a reformulation of $Q'$ on $P_1$. We define the difference $Q - Q''$ as the "semantic loss" of the original query $Q$ when posed to $P_2$.

Our goal now is to track semantic losses in terms of comparing two queries, the original one $Q$, and the reversed one $Q''$. There exist two ways of comparison. The first one is to compare the queries syntactically.

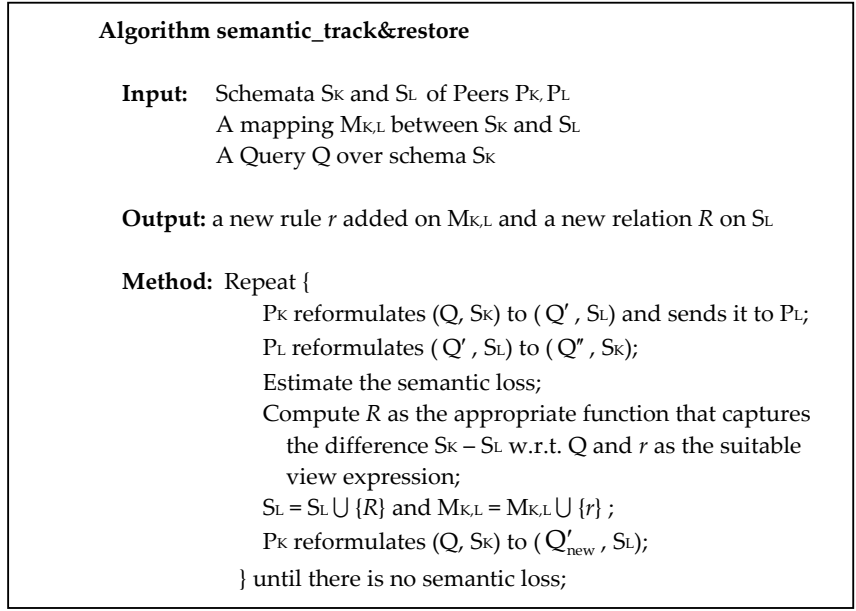

**Figure 2.** An algorithm for tracking and replacing semantic loss in a PDMS network.

This means that we compare the way queries are written (i.e., we compare the "Select", the "From" clauses etc.). From the syntactic differences we can estimate the semantic loss. A second way to compare $Q$ and $Q''$ is to check their results. In this paper we adopt the first way of comparison, since the second cannot always guarantee that identical results of $Q$ and $Q''$ imply absence of loss in $Q'$. Of course given the richness of expression in SQL queries, syntactic comparisons will not always promise the best results. For more general cases one has to use transformations of queries into a canonical representation. But since our main purpose here is to bring out the problem of semantic loss, we deal with the before mentioned simpler case of syntactic comparison.

Suppose now that we have located a loss of semantics happening while query $Q$ propagates from $P_1$ to $P_2$. We need to replace this loss in order for $Q$ to be passed over the next peers. Since the schema of $P_2$ produces the loss a natural idea to overcome this is to embed the loss in a form of a new rule in the mapping that connects the schemas of $P_1$ and $P_2$. Another idea would be to represent the loss within the query, but we choose the previous approach with the perspective that we can circumvent to repeat the process in the case the same or a similar query is posed again.

The algorithm of Figure 2 describes the whole process in a number of steps.

Let's see how the above algorithm works for the motivating example of section 1.2. The algorithm gets as input the schemas $S_{9DC}$ and $S_H$ of the corresponding peers, the mapping $M_{9DC,H}$ and the query $Q_1$ as stated earlier. It reformulates $Q_1$ to $Q_1'$, and again reformulates $Q_1'$ back to $S_{9DC}$ as:

$Q_1''$ : Select PID, skill
From SkilledPerson
Where skill = "Doctor" OR skill = "EMT"

Now since the algorithm has two versions ($Q_1$ and $Q_1''$) of the same query over the same schema it compares them syntactically to estimate the semantic loss. The algorithm sees that these two versions of the same query differ in the "where" clause. In the next step the algorithm calculates the relation $R$ to be added to $S_H$ in order to eliminate the semantic loss (this new relation is only virtual, which means that we don't add new data to node H but we only make an internal change to the schema not visible to the user). For the example such a relation would be a view of the form $CO_{Doctor+Emt}$(SID, skill) over H and a mapping $r$ of the form:

H : $CO_{Doctor+Emt}$ (SID, skill) :-
9DC : SkilledPerson(PID, skill), skill ≠ "Doctor",
skill ≠ "EMT"

We can see that under this setting if $Q_1$ is reformulated over $S_H$ to $Q'_{1,\text{new}}$ we would have:

$Q'_{1,\text{new}}$ : Select SID, "Doctor"
From Doctor
UNION
Select SID, "EMT"
From EMT
UNION
Select SID, skill
From $CO_{\text{Doctor+Emt}}$

And finally if $Q'_{1,\text{new}}$ is reformulated to $Q''_{1,\text{new}}$, we obtain:

$Q''_{1,\text{new}}$ : Select PID, skill
From SkilledPerson
Where skill = "Doctor" OR skill = "EMT"
UNION
Select PID, skill
From SkilledPerson
Where skill ≠ "Doctor" and skill ≠ "EMT"

From the above we can see that $Q_1$ and $Q''_{1,\text{new}}$ coincide semantically which means that the semantic loss is disappeared.

## 2.2. Refining the PDMS network after a semantic loss recovery

Having recovered the semantic loss we are now facing the problem of how to manage the other peers, i.e. how to map the new relation $R$ to their schemas. We consider the idea of exploiting the existing mappings. We compute the transitive relations (i.e., we compare the relations' names and attributes) and find an initial set of correspondences for each neighbor of $P_2$ (the peer that caused the semantic loss). If the resulting set has only one element, then we can easily derive the new rule. Otherwise, we have to reduce the previous set by performing automatic schema matching [5] to a single correspondence which we embed in the form of a rule in the mapping of each neighbor of $P_2$.

Let us see how this idea works in the setting of our motivating example. Since the peer that caused the semantic loss was H, we have to deal with its neighbors FH, LH, and FS. Let's take for example LH. The view we added in $S_H$ is in fact a view of the relation SkilledPerson(PID, skill) of peer 9DC. Computing the transitive relations of SkilledPerson(PID, skill) with the schema $S_{LH}$ we get the set {Staff(SID,…), Schedule(SID,…), InAmbulance(SID, vid)}.

Performing schema matching of this set with $R$ we get for a match the relation Staff(SID,…). This implies that the rule that would link $R$ with $S_{LH}$ is going to be:

H : $CO_{\text{Doctor+Emt}}$ (SID, skill) :-
LH : Staff(SID, firstn, lastn, class), class ≠ "Doctor", class ≠ "EMT"

If we now apply the algorithm of section 2.1 between H and LH we have the following. The algorithm gets as input $S_H$, $S_{LH}$, $M_{H,LH}$ and query $Q_2$:

$Q_2$: Select SID, "Doctor"
From Doctor
UNION
Select SID, "EMT"
From EMT
UNION
Select SID, skill
From $CO_{\text{Doctor+Emt}}$

Then, $Q_2$ is reformulated to LH as $Q'_2$:

$Q'_2$: Select SID, "Doctor"
From Staff, Schedule
Where class = "Doctor" and Staff.SID = Schedule.SID
UNION
Select SID, "EMT"
From Staff, Schedule, InAmbulance
Where class = "EMT" and Staff.SID = Schedule.SID
and Staff.SID = InAmbulance.SID
UNION
Select SID, class
From Staff
Where class ≠ "Doctor" and class ≠ "EMT"

The reformulated query from LH to H is the following:

$Q''_2$ : Select SID, "Doctor"
From Doctor
UNION
Select SID, "EMT"
From EMT
UNION
Select SID, skill
From $CO_{\text{Doctor+Emt}}$

Since the semantic loss $Q_2 - Q''_2 = \varnothing$, we conclude that there is no semantic loss between H and LH which means that from H to LH the query can be propagated without any loss.

## 3. Evaluation and Preliminary results

The lack of existing PDMSs to test on prevents us from fully evaluating our algorithm. Hence our main purpose is to build a small PDMS (like the one in Figure 1) to test our hypothesis. There is an ongoing project in our research team using JXTA [7] that implements a small PDMS topology as a test bed. We defined a set of database schemas and a set of mappings among them. We then populate these databases using a synthetic data generator. We are now in the process of implementing a mechanism (agent) that reformulates the queries taking into consideration the created semantic loss. The agent manages this loss using the algorithm described in Figure 2. Preliminary results are encouraging showing the soundness of our approach. When processing the reformulated queries after the loss is handled, we observe noticeable improvements in the number and the quality of answers.

## 4. Conclusions and Future Work

In this paper, we addressed the problem of semantic loss that affects the propagation of queries in a PDMS network. We proposed an algorithm that employs query comparisons to track such loss. Furthermore, the same algorithm generates the appropriate mapping rules needed for recovering. There is an ongoing project of implementing an agent in JXTA that will evaluate the aforementioned algorithm. Our preliminary results show the soundness of our approach.

As future work, we plan to extend our approach to handle data defined in XML Schema as well as models with rich semantics (ontologies).